\documentclass{amsart}
\usepackage{amssymb}
\begin{document}
\def\Pia{\Pi}
\def\beq{\begin{eqnarray}}
\def\eeq{\end{eqnarray}}
\def\bea{\begin{array}}
\def\eea{\end{array}}
\newcommand{\nn}{\nonumber}
\def\Om{\Omega}
\newcommand{\reals}{\mbox{${\rm I\!R }$}}
\newcommand{\nats}{\mbox{${\rm I\!N }$}}
\def\BB{{\mathcal{B}}}
\def\EE{{\mathcal{E}}}
\def\TTT{{\mathbb{T}}}
\def\vol{\operatorname{vol}}
\def\dvol{d\nu}
\def\text#1{{\hbox{#1}}}
\def\operatorname#1{{\rm#1\,}}
\def\qedbox{\hbox{$\rlap{$\sqcap$}\sqcup$}}
\newtheorem{theorem}{Theorem}
\newtheorem{corollary}[theorem]{Corollary}
\newtheorem{lemma}[theorem]{Lemma}
\newtheorem{remark}[theorem]{Remark}
\newtheorem{ansatz}[theorem]{Ansatz}
\title[Heat content asymptotics]{Heat content asymptotics for\\ spectral boundary conditions}
\author{Peter Gilkey${}^1$, Klaus Kirsten${}^2$, and Jeong-Hyeong Park${}^3$}
\thanks{${}^1$Research partially supported by the NSF 
(USA) and MPI (Leipzig)}
\thanks{${}^2$Research supported by the  MPI (Leipzig)}
\thanks{${}^3$This work was supported by Korea Research 
Foundation Grant (KRF-2000-015-DS0003)}
\begin{address}{PG: Mathematics Department, University of  Oregon, Eugene Or 97403 USA.
\newline Email:gilkey@darkwing.uoregon.edu}\end{address}
\begin{address}{KK:Max Planck Institute for Mathematics in 
the Sciences, Inselstrasse 22-26, 04103 Leipzig, Germany.\quad email:
kirsten@mis.mpg.de}
\end{address}
\begin{address}{JP: Department of Computer \& Applied Mathematics, Honam 
University, Seobongdong 59, Gwangsangu, \newline Gwangju, 506-714 South Korea.
Email:jhpark@honam.honam.ac.kr}\end{address}
\begin{abstract} We study the short time heat 
content asymptotics for spectral boundary conditions. The heat content
coefficients
are shown to be non-local and some preliminary results concerning the structure of the first few
terms are given.\end{abstract}
\maketitle
\section{Introduction}\label{Sect1}
The heat trace asymptotics for `exotic` boundary conditions 
have recently attracted considerable interest; for a survey over this  field
see \cite{gilk02-104-63}. This article is devoted to the analogous
questions for the heat content asymptotics. Whereas some formulas are available for the heat trace asymptotics
with spectral boundary conditions
\cite{dowk99-242-107,gilk00u}, 
nothing is known about the heat content asymptotics in this setting. We shall {\bf postulate} in equation
(\ref{Eqn2}) the existence of an appropriate asymptotic series and then use special case calculations of
spinors on the unit ball to study the heat content asymptotics defined by non-local spectral boundary
conditions -- these are preliminary results in an ongoing investigation.

Spectral boundary conditions were first introduced in the study of the index theorem for manifolds with
boundary by Atiyah, Patodi, and Singer \cite{atiy75-77-43} who assumed that the structures involved were
product near the boundary. Their work was later extended by Grubb and Seeley
\cite{{grub92-17-2031,grub99-37-45,grub93-317-1123,grub95-121-481,grub96-6-31}} to the general setting.

We briefly establish the notational conventions we shall employ
and refer to \cite{dowk99-242-107,gilk00u} for further details. Let ${\mathcal M}$ be a compact 
$m$-dimensional Riemannian manifold with smooth boundary $\partial
{\mathcal M}$. We suppose given unitary vector bundles $E_i$ over 
${\mathcal M}$ and a first order elliptic complex:
$$P:C^\infty (E_1) \rightarrow C^\infty (E_2).$$

As such an elliptic complex need  not admit local boundary
conditions, it is natural to impose non-local spectral boundary conditions, which may be described as follows.
Let $\gamma$ be the leading symbol of $P$. Let $\nabla$ be an auxiliary unitary connection with
$\nabla\gamma=0$; in many applications there is a natural choice available, but it is convenient to work quite
generally for the moment. Let $\nabla_m$ denote covariant differentiation with respect to the inward geodesic
unit normal. Near the boundary, we decompose
$$P=\gamma_m\{\nabla_m+B\},$$
where $B$ is the {\it associated tangential first order operator} on $C^\infty (E_1 |_{\partial
{\mathcal M}})$. We also suppose given an auxiliary self-adjoint endomorphism $\Theta$ of 
$E_1 |_{\partial
{\mathcal M}}$ which we use to define a self-adjoint tangential operator $A$ on $C^\infty (E_1 |_{\partial
{\mathcal M}})$ by setting:
$$A= \textstyle\frac12\{B+B^*\} +\Theta;$$
here the adjoint of $B$ is taken with respect to the structures
on the boundary. The endomorphism $\Theta$ arises naturally in the study of the signature and spin complexes
where the metric is not product near the boundary and in this context is expressible in terms of the second
fundamental form \cite{gilk93-102-129}; $\Theta$ compensates for non-canonical choice of
$\nabla_m$ and plays an important role in the analysis in Section \ref{Sect3}.

Let $\Pia $ be orthogonal projection on the span of the eigenspaces corresponding to the non-negative
eigenvalues of $A$. Let
$P_\Pia$ be the realization of 
$P$ with the boundary condition $\Pia$. The index theorem for manifolds with boundary expresses
$\text{index}(P_\Pia)$ in terms of characteristic forms integrated over $\mathcal{M}$, a compensating
integral involving the second fundamental form over $\partial{\mathcal{M}}$ , and a non-local term (the eta
invariant) for the classic elliptic complexes \cite{atiy75-77-43,gilk93-102-129}.

To simplify the discussion, we shall assume that $E_1=E_2=E$, that $P$ is formally self-adjoint, and that $P$
is of {\it Dirac type}; thus $D:=P^2$ is formally self-adjoint and of {\it Laplace
type}. The leading symbol
$\gamma$ of $P$ is skew-adjoint and 
$$\gamma(\xi)^2=-|\xi|^2\operatorname{Id}\qquad\text{for any}\qquad\xi\in T^*M.$$

$D_\BB$ is the realization of $D$ defined by the operator
$$\BB\psi:=\Pia \psi|_{\partial{\mathcal M}}\oplus\Pia P\psi|_{\partial{\mathcal
M}}.$$ To avoid some technical fuss with the zero mode spectrum, we shall
suppose that $\ker\{A\}=\{0\}$.  We shall also assume that
\begin{equation}\gamma_mA=-A\gamma_m\qquad\text{so}\qquad\gamma_m\Pia=\{\operatorname{Id}-\Pia\}\gamma_m.
\label{Eqn1}\end{equation}
The operators $P_\Pia$ and $D_{\BB}=P_\Pia^2$ are then self-adjoint.

We now describe the fundamental solution of the heat equation. Let $f_1$ represent the initial
``temperature'' distribution of the manifold 
${\mathcal M}$.
The subsequent temperature distribution $h(x,t)$ for $t>0$ is then given as the unique solution of
the equations:
$$\partial_th(x,t)=Dh(x,t),\quad \BB h(x,t)=0,\quad\text{and}\quad\lim_{t\rightarrow0}h(x,t)=f_1(x).$$
Let $\langle\cdot,\cdot\rangle$ denote the Hermitian inner product on $E$. Let
$dx$ and $dy$ be the Riemannian measures 
on $\mathcal{M}$ and $\partial{\mathcal{M}}$, respectively.  Let
$f_2$ represent the ``specific heat'' of the manifold. The total {\it heat energy content} of the manifold is
given by:
$$\beta (f_1, f_2,D,\BB)(t) = \textstyle\textstyle\int_{{\mathcal M}}
dx \,\, \langle h(x,t),f_2 (x)\rangle.$$

It is convenient at this point to review the situation for `standard' boundary conditions. Let $D$ be a
formally self-adjoint second order operator 
of Laplace type whose realization is defined by Dirichlet or Robin boundary
conditions $\BB$. It is known that there is a complete short time asymptotic series of the form:
$$\beta (f_1, f_2, D, \BB ) \sim\sum_{n=0} ^\infty \beta_n 
  (f_1 , f_2,D,\BB) t^{n/2}.$$
The coefficients $\beta_n$ are called the {\it heat content asymptotics}. 

Let Roman
indices $\{i,j,k,l\}$ (resp. $\{a,b\}$) range from $1$ to $m$ (resp. $m-1$) and index an orthonormal
frame for the tangent bundle
$T\mathcal{M}$ (resp. $T\partial\mathcal{M}$). Let
$L_{ab}$ be the components of the second fundamental form and let $R_{ijkl}$ be the components of the Riemann
curvature tensor (with the sign convention $R_{1221}=+1$ on $S^2\subset\mathbb{R}^3$). Let `;' be multiple
covariant differentiation with respect to the Levi-Civita connection on
$\mathcal{M}$ and the natural connection defined by $D$, and let $\mathcal{E}$ be the endomorphism of $E$
defined by
$D$ \cite{berg94-120-48}. We adopt the
Einstein convention and sum over repeated indices. We refer to
\cite{berg93-2-147,berg94-120-48} for the proof of the following Lemma:
\begin{lemma}\label{Lemma1} Let $D$ be an operator of Laplace type on a compact Riemannian manifold
$\mathcal{M}$ with smooth boundary $\partial{\mathcal{M}}$.
\begin{enumerate}
\item If $\BB\psi=\psi|_{\partial\mathcal{M}}$ defines Dirichlet boundary conditions, then
\begin{enumerate}
\item $\beta_0(f_1,f_2,D,\BB)=\textstyle\textstyle\int_{\mathcal{M}}dx \,\, \langle f_1,f_2\rangle$.
\item $\beta_1(f_1,f_2,D,\BB)=-2 \pi^{ -1/2} {\textstyle\int}_{
\partial\mathcal{M}}dy \,\,  \langle f_1,f_2 \rangle$.
\item
$\beta_2(f_1,f_2,D,\BB)=-\textstyle\int_{\mathcal{M}}dx \,\,  \langle Df_1,f_2\rangle
$\newline$+\textstyle\int_{
\partial\mathcal{M}} dy \,\,  \{  \langle  \textstyle\frac12L_{aa} f_1,f_2\rangle - \langle f_1,f_{2;m}
\rangle 
\}$
\item
$\beta_3(f_1,f_2,D,\BB)=-2\pi^{ -1/2}\textstyle\int_{\mathcal{M}}dy \,\,  
\{\frac23\langle f_{1;mm},f_2\rangle+\frac23\langle f_1,f_{2;mm}\rangle
$\newline $-\textstyle \langle f_{1;a},f_{2;a}\rangle
+\langle\mathcal{E}f_,f_2\rangle-\frac23L_{aa}\langle f_{1;m},f_2
\rangle
-\textstyle\frac23L_{aa}\langle f_1,f_{2;m}\rangle
$\newline$+\textstyle\langle (\frac1{12}L_{aa}L_{bb}
-\textstyle\frac16L_{ab}L_{ ab}$
$+\frac16R_{amam})f_1,f_2\rangle\}$.
\end{enumerate}
\item If $\BB\psi=(\nabla_m+S)\psi|_{\partial{\mathcal{M}}}$ defines Robin boundary conditions, then:
\begin{enumerate}
\item $\beta_0(f_1,f_2,D,\BB)=\textstyle\textstyle\int_{\mathcal{M}}dx \,\, \langle f_1,f_2\rangle$.
\item $\beta_1(f_1,f_2,D,\BB)=0$.
\item $\beta_2(f_1,f_2,D,\BB)=-\textstyle\int_{\mathcal{M}}dx \,\, \langle Df_1,f_2\rangle+
\int_{\partial\mathcal{M}}dy \,\,  \langle\BB f_1,f_2\rangle$.
\item
$\beta_3(f_1,f_2,D,\BB)=\textstyle\frac43\cdot\pi^{-1/2}\textstyle\int_{\partial\mathcal{M}}dy \,\,  
\langle\BB f_1,\BB f_2\rangle$.
\end{enumerate}
\end{enumerate}
\end{lemma}

Lemma \ref{Lemma1} shows that the heat content coefficients $\beta_n$ for $n\le 3$ are
locally computable for Dirichlet and Robin boundary conditions. In fact, all the coefficients
$\beta_n$ are locally computable for these boundary conditions. These invariants have been studied
extensively
\cite{berg93-2-147,berg94-120-48,desj94-215-251,avit92-9-1983,avit93-26-823}; we refer to
\cite{gilk02-104-185} for a recent survey article. 

\medbreak We now return to the setting of spectral boundary conditions. We shall {\bf assume} the existence of
a similar asymptotic series for the realization of $D=P^2$ defined by spectral boundary conditions as
$t\rightarrow0$:
\begin{equation}\beta(f_1,f_2,D,\BB)\sim\sum_{n=0}^\infty\beta_n(f_1,f_2,D,\BB)t^{n/2}.\label{Eqn2}
\end{equation}

A natural question then arises, why do we believe there are no log terms -- after all, in the expansion
of the heat trace $\operatorname{Tr}_{L^2}(e^{-tD_\BB})$, log terms appear! 
Our answer is two-fold. First of
all, the heat content asymptotics exhibit various structural simplifications
compared to the heat trace asymptotics. One example is, as we shall 
see presently in Section
\ref{Sect3}, that in the special case of the 
Dirac operator on the unit ball
the relevant universal constants do not depend on the dimension, as they
did for the heat trace aymptotics. Another example is, that the heat content
asymptotics do not show any signs of the loss of strong ellipticity
for the case of oblique boundary conditions \cite{gilk02-59-269}. 
Furthermore, the heat trace asymptotics give rise to log
terms above the dimension of the manifold in the series; 
for $n<m$, the asymptotic coefficients do not have
log terms. So even if the 
situation is as for the heat trace, our results will 
still hold true for $n<m$. 
But this is certainly a question that merits further investigation. 

The first coefficient $\beta_0$ is easily described. Since $\lim_{t\rightarrow0}h(x,t)=f_1(x)$, we have
$\lim _{t\to 0} \beta (f_1,f_2,D,\BB)(t)= \textstyle\int_{{\mathcal M}}
dx \,\,  \langle f_1, f_2\rangle$ and thus, as for Dirichlet and Neumann boundary conditions,
\begin{equation}
\beta_0(f_1,f_2,D,\BB)=\textstyle\textstyle\int_{\mathcal{M}}dx \,\, \langle f_1,f_2\rangle.
\label{Eqn3}\end{equation}

Here is a brief outline to this paper. In Section \ref{Sect2}, we discuss functorial
properties of these invariants, show they are non-local, and outline what we believe the
formula for
$\beta_1$ and $\beta_2$ to be. These results are based on the special case computations in Section \ref{Sect3}
giving a complete calculation of the heat content function for the Dirac operator on the unit ball
with the standard metric.

\section{Functorial Properties}\label{Sect2}

The invariants $\beta_n$ for Dirichlet and Robin boundary conditions have
a number of functorial properties
\cite{berg93-2-147} which extend immediately
to this setting:

\begin{lemma}\label{Lemma2} Let $D_\BB=P_\Pia^2$ be a self-adjoint operator of Laplace type defined by
spectral boundary conditions as described above. Then:
\begin{enumerate}
\item We have $\beta_n(f_1,f_2,c^{-2}D,\BB)=c^{m-n}\beta_n(f_1,f_2,D,\BB)$ for
any $c\in\mathbb{R}^+$.
\item We have
$\beta_n(f_1,f_2,D,\BB)=\beta_n(f_2,f_1,D,\BB)$.
\item If $\BB f_1=0$, then $\beta_n(f_1,f_2,D,\BB)=
-\frac{2} n \beta_{n-2}(Df_1,f_2,D,\BB)$.
\end{enumerate}
\end{lemma}

\begin{proof} The operator $D_\BB$ has a discrete spectral resolution
$\mathcal{S}_{D,\BB}:=\{\psi_\nu,\lambda_\nu\}$
\cite{grub92-17-2031} with associated Fourier coefficients:
$\sigma_\nu(\psi):=\textstyle\int_{\mathcal{M}}dx \,\, \langle\psi,\psi_\nu\rangle$. We may then express:
\begin{eqnarray}
&&h(x,t)=\textstyle\sum_\nu e^{-t\lambda_\nu}\sigma_\nu(f_1)\psi_\nu(x)\qquad\text{so}\label{Eqn4}\\
&&\beta(f_1,f_2,D,\BB)(t)=\textstyle\sum_\nu e^{-t\lambda_\nu}\sigma_\nu(f_1)\sigma_\nu(f_2).
\nn\end{eqnarray}
The estimates of \cite{grub92-17-2031} show these series converge uniformly.

We take
$\Theta(c):=c^{-1}\Theta$ to ensure the associated boundary condition is unchanged. Note that the Riemannian
measure defined by the operator $c^{-2}D$ is $c^mdx$. Since
$\mathcal{S}_{c^{-2}D,\BB}=\{c^{-m/2}\psi_\mu,c^{-2}\lambda_\nu\}$ and
$\sigma_\nu^c(f)=c^{m/2}\sigma_\nu(f)$, we have:
$$\beta(f_1,f_2,c^{-2}D,\BB)(t)=c^m\beta(f_1,f_2,D,\BB)(c^{-2}t).$$
Assertion (1) follows by equating powers of $t$ in this equation.
Since the roles of $f_1$ and $f_2$ are symmetric in display (\ref{Eqn4}), assertion (2) follows. If $\BB
f_1=0$, then the boundary terms vanish and we can integrate by 
parts to compute:
\begin{eqnarray*}
&&\sigma_\nu(Df_1)=\textstyle\int_{\mathcal{M}}dx \,\, \langle 
Df_1,\psi_\nu \rangle
=\textstyle\int_{\mathcal{M}}dx \,\, \langle
f_1,D\psi_\nu \rangle =\lambda_\nu\sigma_\nu(f_1)\qquad\text{so}
\\
&&-\partial_t\beta(f_1,f_2,D,\BB)(t)
    =\textstyle\sum_\nu\lambda_\nu e^{-t\lambda_\nu}\sigma_\nu(f_1)\sigma_\nu(f_2)
=\beta(Df_1,f_2,D,\BB)(t).\end{eqnarray*}
We can now establish (3) by equating terms in the asymptotic expansions for
$\partial_t\beta(f_1,f_2,D,\BB)$ and
$\beta(Df_1,f_2,D,\BB)$.
\end{proof}

The following is an important observation.

\begin{lemma}\label{Lemma3} The heat content coefficients for spectral boundary conditions are not locally
computable
\end{lemma}

\begin{proof} If $\BB f_1 =0$ then $\beta _1 (f_1, f_2,D,\BB)=0$ by Lemma \ref{Lemma2}.
If $\beta_1$ is locally computable, then dimensional analysis (i.e. the scaling property
given by assertion (1) of Lemma \ref{Lemma2}) implies that there must exist a universal
constant $c_0(m)$ so that:
\beq
\beta _1 (f_1, f_2,D,\BB)= 2\pi^{-1/2} \textstyle\int_{\partial {\mathcal M}}
   dy \,\, c_0(m) \,\, \langle f_1 , f_2\rangle .\nn
\eeq
Since generically there are eigensections with $\BB f_1 =0$ but $f_1 |_{\partial {\mathcal M}}
\neq 0$, we must have $c_0 =0$; so far, the argument is exactly the same as for Robin boundary conditions
given in
\cite{desj94-215-251} to prove Lemma \ref{Lemma1} (2b). However, the calculation on the ball that we shall
present in Section \ref{Sect3} shows the power
$t^{1/2}$ is present in the asymptotic expansion with spectral boundary conditions. This contradiction
establishes the Lemma.
\end{proof}

The coefficient $\beta_0$ is
given by equation (\ref{Eqn3}). Using the principle of {\it not feeling the boundary}, we see that the interior integrals defining $\beta_n$
for spectral boundary conditions are the same as those defining $\beta_n$ for either Dirichlet or Robin
boundary conditions. Writing the interior term for $\beta_2$ in the form
$\langle Df_1,f_2\rangle$ destroys the symmetry of Lemma \ref{Lemma2} (2) so instead we use $\langle
Pf_1,Pf_2\rangle$ and add suitable boundary correction terms. The boundary operator $\Pia$ is a 0-th order
operator; it is unaffected by rescaling. To ensure that properties (1) and (3) of Lemma \ref{Lemma2} are
satisfied, i.e.
\begin{eqnarray*}&&\beta_n(f_1,f_2,c^{-2}D,\BB)=c^{m-n}
\beta_n(f_1,f_2,D,\BB)\text{ and }\\&&
\beta_n(f_1,f_2,D,\BB)=\beta_n(f_2,f_1,D,\BB) \end{eqnarray*} 
we are lead to consider the following ansatz for
$\beta_1$ and $\beta_2$ -- the only non-local terms are introduced by the boundary condition.

\begin{ansatz}\label{Ansatz4} There exist universal constants $c_i=c_i(m)$ so that:
\begin{enumerate}
\item 
$\beta_1 (f_1 , f_2,D,\BB) = 2\pi ^{-1/2} \textstyle\int_{\partial {\mathcal M}}
   dy \,\, c_0 (m)\,\, \langle\Pi  f_1,\Pi f_2\rangle$.
\item 
$\beta_2 (f_1 , f_2,D,\BB)= -\textstyle\int_{{\mathcal M}} dx \,\, 
 \langle P f_1 , P f_2\rangle
+\textstyle\int_{\partial {\mathcal M}} dy \,\, 
   \big\{ c_1(m)\langle\Pi \gamma_m P f_1 , \Pi f_2\rangle$
\smallbreak $+c_1(m)\langle\Pi  f_1 , \Pi \gamma_m Pf_2\rangle
+ c_2(m) L_{aa} \langle\Pi  f_1 ,  \Pi f_2\rangle 
   +c_3(m)\langle\Theta\Pi f_1,\Pi f_2\rangle
\big\}.$
\end{enumerate}
\end{ansatz}

The remainder of this note is devoted to the evaluation of these coefficients:

\begin{lemma}\label{Lemma5} We have $c_0(m)=-1$, $c_1(m)=1$, $c_2(m)=\frac12$, and $c_3(m)=0$.
\end{lemma}

It is interesting that these coefficients seem to be dimension free; 
the usual trick of dimension shifting
employed in \cite{berg94-120-48} does not work with spectral boundary conditions. By contrast, the
coefficients in the heat trace asymptotics for spectral boundary conditions are highly dimension dependent
\cite{dowk99-242-107,gilk00u}.

\begin{proof} 
We must ensure the properties of Lemma 2
are satisfied. In particular $\beta_1(f_1,f_2,D,\mathcal{B})$ must vanish if
$\Pi f_1$=0. The boundary term for
$\beta_1$ must be homogeneous of degree $0$. Since $\Pi$ is a $0^{th}$ order
operator,
$\langle\Pi f_1,\Pi f_2\rangle$ has the correct homogeneity, is symmetric in
the roles of $\{f_1,f_2\}$, and vanishes when $\mathcal{B}f_1=0$. This
motivates the formula given in (1).

Consider $\beta_2 (f_1, f_2, D, \BB).$ 
To preserve the interior symmetry, we use the interior integrand
$-\langle Pf_1,$ $Pf_2\rangle$ rather than $-\langle
Df_1,f_2\rangle$. The corresponding integrals are related by the
formula:
\begin{eqnarray*}\textstyle-\int_{\mathcal{M}} dx\langle Df_1,f_2\rangle&=&
\textstyle-\int_{\mathcal{M}} dx\langle
Pf_1,Pf_2\rangle+\int_{\partial\mathcal{M}} dy \langle
\gamma_mPf_1,f_2\rangle .
\end{eqnarray*}
If $\BB f_1=0$, we know 
from Lemma \ref{Lemma2} (3) that 
$$\textstyle \beta_2 (f_1,f_2,D,\BB ) = -\beta_0 (Df_1, f_2, D , \BB).$$
Since $\Pi f_1 = \Pi P f_1 =0$ on $\partial \mathcal{M}$, we use equation
(\ref{Eqn1}) to see that
$$\gamma_m P f_1=\gamma_m(1-\Pi)P f_1=\Pi\gamma_m P f_1\text{ on }\partial M.$$
Consequently, we find:
\beq
\textstyle
-\beta_0 (D f_1, f_2, D, \BB) &=&\textstyle -
\int_{\mathcal{M}}dx \,\, \langle
Pf_1,Pf_2\rangle+\int_{\partial\mathcal{M}}dy \,\,  \langle\Pi\gamma_m P f_1,
f_2\rangle \nn\\
 &=&\textstyle -
\int_{\mathcal{M}}dx \,\, \langle
Pf_1,Pf_2\rangle+\int_{\partial\mathcal{M}}dy \,\,  \langle\Pi\gamma_m P f_1,
 \Pi f_2\rangle .
\eeq 
This shows $c_1 (m) =1$. Lemma 
\ref{Lemma2} (2) shows we need to include $c_1(m)
\langle \Pi f_1 , \Pi \gamma_m P f_2\rangle$ into Ansatz \ref{Ansatz4} (2).
Lemma \ref{Lemma2} (3) shows that apart from the invariants multiplied by 
$c_1 (m)$ additional invariants must disappear if $\Pi f_1 = \Pi P f_1 =0$.
Applying Lemma \ref{Lemma2} (1) 
this allows for the occurrence of the remaining 
terms in Ansatz \ref{Ansatz4} (2).

Replacing $\Theta$ by $\Theta+\varepsilon$ where $\varepsilon$ is a small positive real constant does not
change the spectral projection $\Pi$ and hence does not change
$\beta_n$. Thus $c_3(m)=0$. We postpone the evaluation of the 
remaining constants, $c_0 (m)$ and $c_2(m)$, until Section
\ref{Sect3}.
\end{proof}

\section{Calculations on the Ball}\label{Sect3}

The eigenvalue problem on the ball is known \cite{dowk96-13-2911,gilk00u} and 
we will only summarize the relevant equations for the present context.
We use the following representation of the $\gamma$-matrices projected along
$e_j$:
\begin{eqnarray}
&&\gamma_{a(m)}=\left(
   \begin{array}{cc}
               0 &  \sqrt{-1}\cdot \gamma_{a(m-1)}    \\
      -\sqrt{-1}\cdot \gamma_{a(m-1)}    &     0
    \end{array}    \right)\text{ and }\nn\\
\quad
&&\gamma_{m(m)} = \left(
     \begin{array}{cc}
         0       &    \sqrt{-1}\cdot 1_{m-1}   \\
    \sqrt{-1}\cdot 1_{m-1}\quad\    &      0
    \end{array}   \right)   .\nonumber
\end{eqnarray}
We decompose $\nabla_j = e_j + \omega_j$ where
$\omega_j=\frac 1 4 \Gamma_{jkl} \gamma_{k(m)}
\gamma_{l(m)}$ is the connection $1$ form
of the spin connection -- i.e.
$$
\nabla_a = \frac 1 r\left( \left(
        \begin{array}{cc}
        \tilde{\nabla}_a & 0 \\
         0 & \tilde{\nabla}_a
         \end{array}  \right)  +\frac 1 2 \gamma_{m(m)}^{-1} 
    \gamma_{a(m)}\right).
$$
Let $P$ and $\bar P$ be the Dirac operator on the ball and the sphere
respectively. 
In the notation established above, the Dirac operator on the ball is
\begin{eqnarray}
&&P=\left(\frac{\partial}{\partial x_m}-\frac{m-1}{2r} \right) \gamma_{m(m)}
         +\frac 1 r \left(
\begin{array}{cc}
       0   & \sqrt{-1} \bar P \\
    -\sqrt{-1} \bar P  & 0
\end{array}   \right).\nn
\end{eqnarray}
Let $\varphi_\pm$ and ${\mathcal Z}_\pm^{(n)}$ denote the eigen functions
of $P$ and $\bar P$ respectively,
\beq
&&P \varphi_\pm= \pm \mu \varphi_\pm ,  \,\,
\bar P {\mathcal Z}_\pm ^{(n)}= \pm \left( n+\frac{m-1} 2 \right) 
     {\mathcal Z}_\pm^{(n)} \text{ for }n=0,1,2,... \nn
\eeq
A complete set of eigen functions is
\begin{eqnarray*}
\varphi_{\pm}^{(+)}&=&{\frac{C}{r^{(m-2)/2}}} \left(
     \begin{array}{c}
        iJ_{n+m/2}(\mu r)
       \,Z^{(n)}_+(\Omega )  \\
     \pm J_{n+m/2-1}(\mu r)\,Z^{(n)}_+(\Omega )
        \end{array}  \right) , \text{ and} \\
\varphi_{\pm}^{(-)}&=&{\frac{C}{r^{(m-2)/2}}}\left(
     \begin{array}{c}
     \pm J_{n+m/2-1}(\mu r)\,Z^{(n)}_
-(\Omega )  \\
   iJ_{n+m/2}(\mu r)\,Z^{(n)}_-(\Omega ) \end{array}
\right),\text{where}\\
C&=&{J_{n+m/2} (\mu)}^{-1} \nonumber
\end{eqnarray*}
is the radial normalization constant.
With the choice $\Theta = (m-1)/2\,\, 1_m$, the boundary operator 
$A$ used to define spectral boundary conditions then reads
\begin{eqnarray}
A=\left(
\begin{array}{cc}
-\bar P & 0 \\
0 & \bar P
\end{array} \right). \nn
\end{eqnarray}
It is easy to determine the discrete spectral resolution of $A$. One can show
\beq
A\left(\begin{array}{c}
          {\mathcal Z}_+^{(n)}(\Omega ) \\
            {\mathcal Z}_-^{(n)} ( \Omega )
        \end{array} \right) &=& -\left( n+\frac{m-1} 2 \right)
\left(\bea {c}
          {\mathcal Z}_+^{(n)} (\Omega )\\
            {\mathcal Z}_-^{(n)} (\Omega )
        \eea \right)\text{ and} \nn\\
A\left(\bea {c}
          {\mathcal Z}_-^{(n)} (\Omega )\\
            {\mathcal Z}_+^{(n)} (\Omega )
        \eea \right) &=&  \left( n+\frac{m-1} 2 \right)
\left(\bea {c}
          {\mathcal Z}_-^{(n)} (\Omega )\\
            {\mathcal Z}_+^{(n)} (\Omega )
        \eea \right)\text{ for }n=0,1,....\nn
\eeq
Thus there is no zero mode spectrum. Spectral boundary conditions suppress the non-negative spectrum of $A$, 
which yields the implicit eigenvalue equation
\beq
J_{n+m/2-1} (\mu ) =0, \quad n=0,1,2,...\nn
\eeq
We will analyze the heat content asymptotics by considering the associated
zeta function. Denoting by $\phi_k$ the full set of eigen functions,
$\phi_k = (\varphi_\pm^{(+)}, \varphi_\pm ^{(-)})$, we write
\beq
\zeta (s,f_1,f_2,D,\BB) = \sum_k \lambda_k^{-s} \,\,
          (f_1, \phi_k ) _{L^2} \,\,
          (\phi_k , f_2 ) _{L^2}.
\nn
\eeq
The asymptotic coefficients $\beta_n$ given in equation (\ref{Eqn2}) are then given by:
\beq
\beta_{2k} (f_1, f_2,D,\BB) &=& \textstyle\frac{(-1)^k}{k!}
         \zeta (-k, f_1, f_2,D,\BB) , \label{Eqn5}\\
\beta_{2k+1}  (f_1, f_2, D, \BB) &=& \Gamma (-k-\textstyle\frac12)
          \mbox{Res } \zeta (-k-\textstyle\frac12,  f_1, f_2,D,\BB)
  \label{Eqn6} . 
\eeq
We now proceed with the explicit 
calculation of the heat content asymptotics on
the ball. The ability to perform a special case calculation strongly 
depends on the choice of the initial temperature $f_1$ and of the specific
heat $f_2$. We establish Lemma \ref{Lemma3} giving the non-locality
of the heat content asymptotics by considering the functions:
$$f_i^{(1)}=f^{(1)}=\left(\begin{matrix}0\\Z_+^{(0)}(\Om)\end{matrix}\right)\quad\text{ and }\quad
  f_i^{(2)}=f^{(2)}=\left(\begin{matrix}rZ_+^{(0)}(\Om)\\0\end{matrix}\right).$$
These spinors have the property $\Pia  f^{(1)} = f^{(1)}$, whereas
$\Pia  f^{(2)} = 0 $.  
Furthermore, 
because the spinor spherical harmonis involved 
are orthogonal,
$$
(f^{(j)}, \varphi^{(-)}_\pm)_{L^2} =0.
$$To evaluate the 
scalar product with $\varphi_\pm ^{(+)}$, the relevant $r$-integrals are
\cite{grad65b}
\beq
\textstyle\int_0^1dx \,\, x^{\nu +1} J_{\nu} (\mu x) = \frac 1 \mu J_{\nu +1} (\mu ).\nn
\eeq
We first proceed with $f^{(1)}$. We have 
\beq
(f^{(1)}, \varphi_\pm ^{(+)})_{L^2} = \textstyle\pm \frac 1 {J_{m/2} (\mu )}
   \textstyle\int_0^1 dr \,\, r^{m/2} J_{m/2 -1} (\mu r) = \pm \frac 1 \mu .\nn
\eeq
Let the contour $\gamma$ enclose all  positive zeroes of $J_{m/2-1} (k)$.
In the contour integral formalism developed in 
\cite{bord96-37-895,bord96-182-371}, we have the following representation:
\beq
\zeta (s , f^{(1)}, f^{(1)},D,\BB) = -2 
\textstyle\int_\gamma \frac{dk}{2\pi i} k^{-2s-2} \frac{\partial}{\partial k} 
    \ln J_{m/2-1} (k) .\nn
\eeq

Deforming the contour towards the imaginary axis we arrive at
\beq
\zeta (s, f^{(1)}, f^{(1)},D,\BB) &=& 2 \textstyle\int_{\gamma_\epsilon} 
\frac{dk}{2\pi i} k^{-2s-2} \frac{\partial}{\partial k} 
    \ln J_{m/2-1} (k) \nn\\
   & & -2\frac{\sin (\pi s)} \pi \textstyle\int_\epsilon^\infty dk \,\, 
   k^{-2s-2} 
   \frac{\partial}{\partial k} 
   \ln I_{m/2-1} (k) ,\nn
\eeq
with $0<\epsilon \in\reals$ smaller than the first positive zero of $J_{m/2
-1}  (k)$ and with a semicircle around zero of radius
$\epsilon$ in the right half plane:
$$\gamma_\epsilon = \{\epsilon e^{it}, t\in
[\pi/2 , -\pi /2]\}.$$

We determine the contributions of the two terms separately. Although
each term depends on $\epsilon$ we know the sum will be independent
of $\epsilon$ and for that reason we concentrate on the 
$\epsilon$-independent
part of both terms. The relevant information to recover the properties 
in (\ref{Eqn5}) and (\ref{Eqn6}) is encoded in the small-$k$ 
behavior of $J_{m/2-1} (k)$. The full expansion is \cite{grad65b}
\beq
J_\nu (k) = \left( \frac k 2 \right) ^ \nu 
   \frac 1 {\Gamma (\nu +1)} \sum_{l=0}^\infty 
       (-1)^l \frac{\Gamma (\nu +1)} {l! \Gamma (\nu +l+1)} 
   \left( \frac k 2 \right) ^{2l} , \nn
\eeq
from which we may obtain the expansion:
\beq
\ln J_\nu (k) = \nu \ln k -\ln \left[ 2^\nu \Gamma (\nu +1) \right] 
 + \sum_{l=1} ^\infty g_l k^{2l}. \nn
\eeq
Hereby the coefficients $g_l$ are defined. In particular
$g_1=-1/[4(\nu +1)]$. It is easy to see that from the small-$\epsilon$
circle no contribution to the residues results, but that 
\beq
\zeta_\epsilon (0, f^{(1)}, f^{(1)}, D, \BB ) =\textstyle\frac1m, \quad 
\zeta_\epsilon (-1, f^{(1)}, f^{(1)}, D, \BB ) = -\frac {m} 2 +1.\nn
\eeq
We consider next the contribution along the imaginary axis. This time the 
large-$k$ behavior is needed to determine the information in equations
(\ref{Eqn5}) and (\ref{Eqn6}). For large $k$, the relevant expansion
of the Bessel function is \cite{grad65b}
\beq
I_\nu (k) \sim \frac{e^k}{\sqrt{2\pi k} } 
 \sum_{l=0}^\infty \frac{(-1)^l}{(2k)^l} \frac{\Gamma (\nu +1/2 +l)}
 {l! \Gamma (\nu +1/2 -l)} \nn
\eeq
and we define coefficients $h_j$ by 
\beq
\ln I_\nu (k) \sim k-\frac 1 2 \ln (2\pi k) + 
  \sum_{j+1} ^\infty h_j k^{-j} .\nn
\eeq
The needed $k$-integrals are trivial,
$
\textstyle\int_\epsilon ^\infty dx \,\,  x^{-\alpha} = \frac{ \epsilon^{1-\alpha}}
 {\alpha -1} ,\nn
$
and the $\epsilon$-independent pieces at the particular values of $s$ needed
are easily obtained. We have that: 
\beq
\zeta (0, f^{(1)}, f^{(1)},D,\BB) &=& 0 , \nn\\
\text{Res }(-{\textstyle\frac12},f^{(1)}, f^{(1)},D,\BB) &=& \textstyle\frac 1 \pi , \nn\\
\zeta (-1, f^{(1)}, f^{(1)},D,\BB) &=& -\textstyle\frac{m-1} 2 , \nn\\
\text{Res } (-k-{\textstyle\frac12},f^{(1)}, f^{(1)},D,\BB) &=&
 \pi^{-1}(-1)^{k+1} (2k-1) h_{2k-1} , \quad k\in \nats, \nn\\
\zeta (-k, f^{(1)}, f^{(1)},D,\BB) &=& (-1)^k 2 (k-1) h_{2k-2} , 
\quad k-1 \in \nats . \nn
\eeq
For the heat content coefficients we conclude from (\ref{Eqn5}) 
and (\ref{Eqn6}) that 
\beq
\beta_0 (f^{(1)}, f^{(1)},D,\BB) &=& \textstyle\frac 1 m , \quad
\beta_1 (f^{(1)}, f^{(1)},D,\BB) =\textstyle -\frac 2 {\sqrt{\pi}} , \nn\\
\beta_2 (f^{(1)}, f^{(1)},D,\BB) &=& \textstyle\frac {m-1} 2 ,\nn
\eeq
which is in agreement with 
\beq
\beta_0 (f^{(1)}, f^{(1)},D,\BB) &=& \textstyle\int_{{\mathcal M}} dx \,\,  
  \langle f^{(1)} ,  f^{(1)}\rangle, \nn\\
\beta_1 (f^{(1)}, f^{(1)},D,\BB) &=& -2 \pi^{-1/2}  \textstyle\int_{\partial
{\mathcal M}} dy \,\,   
 \langle\Pia   f^{(1)} ,  \Pia  f^{(1)}\rangle , \nn\\
\beta_2 (f^{(1)}, f^{(1)},D,\BB) &=& \frac 1 2 \textstyle\int_{\partial
{\mathcal M}} dy \,\,   L_{aa}\langle\Pia  f^{(1)} ,  
   \Pia  f^{(1)}\rangle ,\nn
\eeq
and whereby $c_0 = -1$ and $c_2 = \frac 1 2$ 
has been determined; thus completing the proof of Lemma \ref{Lemma5}. As a further check that it is really
the projection
$\Pia  $ that enters the coefficients 
we next consider the function 
$f^{(2)}$. We have
$
(f^{(2)}, \varphi_\pm^{(+)})_{L^2} = \frac {iJ_{m/2+1}}{kJ_{m/2}}.\nn$
So the starting point for the associated zeta function is 
\beq
\zeta (s, f^{(2)}, f^{(2)},D,\BB) = 2 \textstyle\int_\gamma 
\frac{dk}{2\pi i} k^{-2s-2} \frac{J_{m/2+1} ^2 (k) }
 {J_{m/2} ^2 (k) } \frac{\partial}{\partial k} \ln J_{m/2-1} (k) .
\label{Eqn7}
\eeq
Note that the contour $\gamma$ encloses the zeroes of $J_{m/2-1} (k)$ 
only. In fact, it is possible to place the contour such that no zeroes 
of $J_{m/2} (k)$ are enclosed, because the zeroes of $J_{m/2-1}(k)$ 
are simple, thus $0\neq J'_{m/2-1}  (k) = -J_{m/2} (k)$. 

Next we use \cite{grad65b}
\beq
k J_{m/2+1} (k) &=& m J_{m/2} (k) - k J_{m/2-1} ,\qquad\text{ so }\nn\\
k^2 J_{m/2+1} ^2 (k) &=& m^2 J_{m/2} ^2 (k) + k^2 J_{m/2-1} ^2 -
2 k m J_{m/2} (k) J_{m/2-1} (k) .\nn
\eeq
We use the residue theorem, to see only the first term can contribute
in (\ref{Eqn7}). Thus
\beq
\zeta (s, f^{(2)}, f^{(2)},D,\BB) = 2 m^2 \textstyle\int_\gamma 
\frac{dk}{2\pi i} k^{-2s-4} 
\frac{\partial}{\partial k} \ln J_{m/2-1} (k)  .\nn
\eeq
Proceeding as before, we find
\beq
&&\zeta (0, f^{(2)}, f^{(2)},D,\BB)= \textstyle\frac 1 {m+2},\qquad
\mbox{Res }\zeta (-1/2, f^{(2)}, f^{(2)},D,\BB) = 0 ,\nn\\ 
&&\zeta (-1, f^{(2)}, f^{(2)},D,\BB) = m ,\qquad\text{ so }\quad
\beta_0 ( f^{(2)}, f^{(2)},D,\BB) =\textstyle \frac 1 {m+2} ,\nn\\
&&\beta_1 ( f^{(2)}, f^{(2)},D,\BB) = 0,\qquad\qquad\text{and}\quad
\beta_2 ( f^{(2)}, f^{(2)},D,\BB) = -m .\nn
\eeq
This is consistent the form given in Ansatz \ref{Ansatz4}.
No boundary contributions are found as a result of $\Pia  f^{(2)}
=0$.

\end{document}